\documentclass[11pt]{article}
\usepackage[margin=1in]{geometry}
\usepackage[T1]{fontenc}
\usepackage[utf8]{inputenc}
\usepackage{lmodern,microtype}
\usepackage{amsmath,amssymb,graphicx,subcaption}
\usepackage{xcolor}
\usepackage{tikz}
\usetikzlibrary{arrows.meta}
\usepackage{xurl}
\usepackage[hidelinks]{hyperref}
\usepackage[section]{placeins}

\font\code=cmtt10

\newcommand{\vect}[1]{\boldsymbol{#1}}
\newcommand{\dd}{\,\mathrm{d}}
\title{Three-Dimensional Simulation of Flood Propagation\\
at Syabrubesi, Nepal \thanks{This work is partially supported by the Office of Naval Research under Award No. N00014-24-1-2147, the National
Science Foundation under Grant DMS-2408877, the Air Force Office of Scientific Research under Award No. FA9550-221-0248, and SURE-AI Centre grant 357482, Research Council
of Norway.}}
\author{Rainald L\"ohner\thanks{Center for Computational Fluid Dynamics and
Department of Physics and Astronomy, George Mason University,
Fairfax, Virginia, USA.}
\and 
Harbir Antil\thanks{Department of Mathematical Sciences and Center
for Mathematics and Artificial Intelligence, George Mason University,
Fairfax, Virginia, USA.}}
\date{}
\begin{document}
\maketitle

\begin{abstract}
Three-dimensional flood propagation through the Bhote Koshi--Langtang
Khola junction at Syabrubesi, Nepal, is studied using an incompressible
free-surface model. The calculation is motivated by the August 2026 flood
and examines the passage of an imposed inflow through the local bends
and changes in channel width. The Navier--Stokes equations and an
interface transport equation are solved on unstructured
tetrahedral meshes constructed from pre-event terrain data. For an
inlet speed of $10\,\mathrm{m\,s^{-1}}$, the solution indicates 
surface speeds above $40\,\mathrm{m\,s^{-1}}$. The advancing
front spreads across the junction; as the flow develops, water deepens
and slows around the bend. Higher speeds persist in the confined
upstream and downstream regions.
\end{abstract}
\noindent\textbf{Keywords:} flood propagation; free-surface flow;
Navier--Stokes equations; unstructured finite elements; FEFLO

\section{Introduction}

Flood propagation in steep mountain valleys is affected by channel
constrictions, bends and tributary junctions. These features redirect
the flow and cause substantial variations in velocity and water depth.
At Syabrubesi, Nepal, the Langtang Khola joins the Bhote Koshi near a
pronounced change in the direction of the main channel. The flood of
26 August 2026 passed through this reach after travelling downstream
from a rock--ice failure near the Nepal--China border \cite{Li2026}.

Recent studies have examined the source failure, entrainment and
downstream transformation of the moving material \cite{Li2026,Guo2026}.
Numerical reconstructions have also considered propagation over the
wider river corridor \cite{Park2026,Matsushima2026}. At the local scale,
Weedman \cite{Weedman2026} reports a shallow-water calculation around
Syabrubesi using an $8\,\mathrm{m}$ terrain grid. Such calculations
provide a description of the flood extent and its propagation along
the valley. Here we examine the three-dimensional flow through the
Syabrubesi junction, where the river widens and turns southwest.

A three-dimensional free-surface formulation permits the local motion
to be computed without assuming a hydrostatic pressure distribution or
replacing the velocity field by its depth average. We use the
unstructured finite-element solver FEFLO and the free-surface method
developed in \cite{Loh99,Lohner2006}. The formulation combines a
velocity-prediction and pressure-correction scheme with edge-based
spatial discretization and transport of a scalar interface indicator.
Local mesh refinement allows the river channel to be resolved within
the surrounding terrain.

The computational domain covers a
$2\,\mathrm{km}\times2\,\mathrm{km}$ area constructed from pre-event
elevation data. An inflow is prescribed at the northern river section,
and the moving material is represented by an incompressible liquid with the
density of water. The calculation concerns the passage of this flow
through the junction over a fixed bed. We examine the advance of the
free surface and the accompanying changes in velocity, depth and
Froude number. The numerical formulation is described first, followed
by the computational model and the results.

\section{Mathematical and Numerical Model}
\label{sec:method}

\subsection{Modeling the Mudflow}

For the purposes of this study, the mudflow was assumed to be an incompressible 
turbulent flow, properly described by the Navier-Stokes equations given by:
\begin{align}
 \rho\frac{\partial\vect v}{\partial t}
 +\rho(\vect v\cdot\nabla)\vect v+\nabla p
 &=\nabla \cdot (\mu\nabla \vect v)+\rho\vect g,
 \label{eq:momentum}\\
 \nabla\cdot\vect v&=0.
 \label{eq:continuity}
\end{align}
Here $\rho$ denotes the density, $\vect v$ the velocity, $p$ the pressure, 
$\mu$ the dynamic and turbulent viscosity and $\vect g$ the gravitational
acceleration. We remark that both the gaseous and liquid phases are 
considered incompressible, thus \eqref{eq:continuity}. 
The free surface (liquid-gas interface) is represented by a scalar indicator
$\Phi$ transported according to
\begin{equation}
 \frac{\partial\Phi}{\partial t}+\vect v\cdot\nabla\Phi=0.
 \label{eq:transport}
\end{equation}
For the classic VOF technique, $\Phi$ represents the total density 
of the material in a cell/element or control volume{\cite{Nic75, 
Hir81, Unv92, Sus94, Sca99, Che99, Fek99, Bia04, Hui04}}.
For pseudo-concentration techniques, $\Phi$ represents the 
percentage of liquid in a cell/element
or control volume. For the level set (LS) approach $\Phi$ represents 
the signed distance to the interface{\cite{Sus94, Enr03}}. For a 
combinations of these approaches, see Sussman and 
Puckett{\cite{Sus00}}.

The code chosen to compute the flow is FEFLO. Since the mid 
1990s {\cite{Loh90,Mar92,Ram96,Loh99}} the numerical
schemes used in FEFLO to solve the incompressible Navier-Stokes
equations given by Eqns.~\eqref{eq:momentum}--\eqref{eq:continuity} have 
been based on the following criteria:
\begin{itemize}
\item[-] Spatial discretization using {\bf unstructured grids} 
(in order to allow for arbitrary geometries and adaptive refinement);
\item[-] Spatial approximation of unknowns with {\bf simple finite 
elements} (in order to have a simple input/output and code structure);
\item[-] Temporal approximation using {\bf implicit integration of 
viscous terms and pressure} (the interesting scales are the ones 
associated with advection);
\item[-] Temporal approximation using {\bf explicit integration of 
advective terms};
\item[-] {\bf Low-storage, iterative solvers} for the resulting 
systems of equations (in order to solve large 3-D problems); and
\item[-] Steady results that are {\bf independent from the timestep} 
chosen (in order to have confidence in convergence studies).
\end{itemize}

\subsection{Temporal discretization}

For the flooding cases considered here 
the important physical phenomena propagate with the {\bf advective}
timescales. This implies that the advective terms require
an explicit time integration. Diffusive phenomena typically occur 
at a much faster rate, and can/should
therefore be integrated implicitly. Given that the pressure establishes
itself immediately through the pressure-Poisson equation, an implicit
integration of pressure is also required.
The hyperbolic character of the advection operator and the elliptic
character of the pressure-Poisson equation have led to a number of
so-called projection schemes. The key idea is to predict first a
velocity field from the current flow variables without taking the
divergence constraint into account. In a second step, the divergence
constraint is enforced by solving a pressure-Poisson equation. The
velocity increment can therefore be separated into an
advective-diffusive and pressure increment:
\begin{equation}
 \vect v^{n+1}=\vect v^n+\Delta\vect v^a+\Delta\vect v^p
             =\vect v^*+\Delta\vect v^p.
 \label{eq:split}
\end{equation}
The superscript $n$ denotes the time level, $\Delta t=t_{n+1}-t_n$,
and $\vect v^*$ is the predicted velocity.

For an explicit (forward Euler) integration of the advective terms, 
with implicit integration of the viscous terms, one complete timestep 
is given by:

\begin{itemize}
\item[-] {\bf Advective-Diffusive Prediction}:
$\vect v^n \rightarrow \vect v^{*}$
\begin{equation}
 \left[\frac{\rho}{\Delta t}-\theta\nabla \cdot (\mu\nabla) \right]
 (\vect v^*-\vect v^n)
 +\rho(\vect v^n\cdot\nabla)\vect v^n+\nabla p^n
 =\nabla \cdot (\mu \nabla \vect v^n)+\rho\vect g.
 \label{eq:predictor}
\end{equation}
The factor $\theta$ determines the implicitness of the viscous term:
$\theta=1$ gives backward Euler and $\theta=1/2$ gives the
Crank--Nicolson treatment of this term.

\item[-] {\bf Pressure correction}: 
\begin{align}
 \nabla\cdot\vect v^{n+1}&=0,
 \label{eq:corrected-divergence}\\
 \rho\frac{\vect v^{n+1}-\vect v^*}{\Delta t}
 +\nabla(p^{n+1}-p^n)&=0.
 \label{eq:pressure-correction}
\end{align}
Taking the divergence of Eq.~\eqref{eq:pressure-correction} and using
Eq.~\eqref{eq:corrected-divergence} yields the pressure-Poisson equation
\begin{equation}
 \nabla\cdot\left[\frac{1}{\rho}\nabla(p^{n+1}-p^n)\right]
 =\frac{\nabla\cdot\vect v^*}{\Delta t}.
 \label{eq:poisson}
\end{equation}

\item[-] {\bf Velocity correction}: $\vect v^* \rightarrow \vect v^{n+1}$
\begin{equation}
 \vect v^{n+1}=\vect v^*
 -\frac{\Delta t}{\rho}\nabla(p^{n+1}-p^n).
 \label{eq:velocity-correction}
\end{equation}

\end{itemize}

At steady state, $\vect v^{*}=\vect v^n=\vect v^{n+1}$ and the residuals 
of the pressure correction vanish, implying that the result does not 
depend on the timestep $\Delta t$. $\theta$ denotes the implicitness-factor 
for the viscous terms ($\theta=1$: 1st order, fully implicit, 
$\theta=0.5$: 2nd order, Crank-Nicholson). One can replace the 
one-step explicit advective-diffusive predictor by a multistage 
Runge-Kutta scheme {\cite{Loh04}}, allowing for higher accuracy in 
the advection-dominated regions and larger timesteps 
without a noticeable increment in CPU cost.

For $k$ stages, a time-accurate Runge-Kutta scheme of order $k$ for the 
advective parts may be written as:
\begin{equation}
 \begin{split}
 \rho\vect v^{(i)}=\rho\vect v^n+\alpha^{(i)}\gamma\Delta t
 \bigl[&-\rho(\vect v^{(i-1)}\cdot\nabla)\vect v^{(i-1)}
       -\nabla p^n\\
       &+\nabla \cdot (\mu\nabla \vect v^{(i-1)})+\rho\vect g\bigr],
 \qquad i=1,\ldots,k-1,
 \end{split}
 \label{eq:stages}
\end{equation}
with $\alpha^{(i)}=1/(k+1-i)$. The last stage satisfies:
\begin{equation}
 \begin{split}
 \left[\frac{\rho}{\Delta t}-\theta \nabla \cdot (\mu \nabla) \right]
 (\vect v^{(k)}-\vect v^n)
 +\rho(\vect v^{(k-1)}\cdot\nabla)\vect v^{(k-1)}+\nabla p^n\\
 =\nabla \cdot (\mu\nabla \vect v^{(k-1)})+\rho\vect g,
 \end{split}
 \label{eq:last-stage}
\end{equation}
and $\vect v^*=\vect v^{(k)}$.  As compared to the original scheme 
given by \eqref{eq:predictor}, the $k-1$ stages in \eqref{eq:stages} 
may be seen as a predictor 
(or replacement) of $\vect v^n$ by $\vect v^{k-1}$. The original 
right-hand side has not been modified, so that at steady-state 
$\vect v^n=\vect v^{k-1}$, preserving the requirement that the 
steady-state be independent of the timestep $\Delta t$.

The factor $\gamma$ denotes the local ratio of the stability limit for
explicit timestepping for the viscous terms versus the timestep chosen.
Given that the advective and viscous timestep limits are proportional 
to:
\begin{equation}
 \Delta t_a\sim\frac{h}{|\vect v|},\qquad
 \Delta t_v\sim\frac{\rho h^2}{\mu}.
 \label{eq:time-scales}
\end{equation}
Their ratio is of the order of the element Reynolds number,
\begin{equation}
\gamma = 
 \frac{\Delta t_v}{\Delta t_a}\sim
 \frac{\rho|\vect v|h}{\mu}\sim \mathrm{Re}_h.
 \label{eq:element-reynolds}
\end{equation}
For a time step chosen on the advective scale, the factor is taken as
\begin{equation}
 \gamma=\min(1,\mathrm{Re}_h).
 \label{eq:gamma}
\end{equation}
In regions away from boundary layers, this factor is $O(1)$, 
implying that a high-order Runge-Kutta scheme is recovered. 
Conversely, for regions where $\mathrm{Re}_h=O(0)$, the scheme reverts back 
to the original one (\eqref{eq:predictor}).
Projection schemes of this kind (explicit advection with a variety of
schemes, implicit diffusion, pressure-Poisson equation for either the
pressure or pressure increments) 
have been widely used in conjunction with spatial discretizations
based on finite differences {\cite{Kim85, Bel89, Bel92, Ale96}}, finite 
volumes {\cite{Kal96}},  and finite elements {\cite{Gre82, Don82, Gre90, 
Loh90, Mar92, Ram96, Loh99, Tak01, Eat01, Kar01, Cod01, Li02, 
Kar02, Loh04, Cam04}}. 

\medskip
One {\bf complete timestep} is then comprised of the following 
substeps:
\begin{itemize}
\item[-] Predict velocity (advective-diffusive predictor, 
Eqns. \eqref{eq:predictor}, \eqref{eq:stages}, and \eqref{eq:last-stage}; 
\item[-] Extrapolate the pressure (imposition of boundary conditions);
\item[-] Update the pressure (Eqn.~\eqref{eq:pressure-correction});
\item[-] Correct the velocity field (Eqn.~\eqref{eq:velocity-correction});
\item[-] Extrapolate the velocity field; and 
\item[-] Update the scalar interface indicator.
\end{itemize}

\subsection{Spatial Discretization}

As stated before, we desire a spatial discretization with unstructured
grids in order to:
\begin{itemize}
\item[-] Approximate arbitrary domains, and
\item[-] Perform adaptive refinement in a straightforward manner, i.e.
without changes to the solver.
\end{itemize}
From a numerical point of view, the difficulties in solving 
Eqns.~\eqref{eq:momentum}, \eqref{eq:continuity}, and \eqref{eq:transport} 
are the usual ones. First-order derivatives that are simply discretized via
a Galerkin approximation can lead to overshoots, oscillations or instabilities).
On the other hand, discretizing second-order derivatives by a 
straightforward Galerkin approximation leads to stable schemes. In the sequel,
the advection operator and then the divergence will be treated.
Given that for tetrahedral grids solvers based on edge data 
structures incur a much lower indirect addressing and CPU overhead 
than those based on element data structures {\cite{Loh01}}, only these
will be considered.

\subsubsection{The Advection Operator}

Let $N^i$ denote the shape function at node $i$. 
For each edge joining nodes $i$ and $j$, the edge coefficients are
\begin{equation}
 d_m^{ij}=\frac12\int_\Omega
 \left(\frac{\partial N^i}{\partial x_m}N^j
       -\frac{\partial N^j}{\partial x_m}N^i\right)\dd\Omega,
 \qquad m=1,2,3,
 \label{eq:edge-coefficients}
\end{equation}
and
\begin{equation}
 D^{ij}=\left(\sum_{m=1}^3(d_m^{ij})^2\right)^{1/2},
 \qquad S_m^{ij}=\frac{d_m^{ij}}{D^{ij}}.
 \label{eq:edge-direction}
\end{equation}
Here $\Omega$ is the computational domain and $\vect S^{ij}$ is the
direction defined by the edge coefficients.

For the momentum equation, the interior advective contribution to the 
nodal right-hand side is
\begin{equation}
 \vect r^i=\sum_{j\in\mathcal N(i)}D^{ij}\vect{\mathcal F}_{ij},
 \qquad \vect{\mathcal F}_{ij}=\vect f_i+\vect f_j,
 \label{eq:advective-residual}
\end{equation}
where $\mathcal N(i)$ denotes the neighbouring nodes and
\begin{equation}
 \vect f_i=(\vect S^{ij}\cdot\vect v_i)\vect v_i,
 \qquad
 \vect f_j=(\vect S^{ij}\cdot\vect v_j)\vect v_j.
 \label{eq:edge-flux}
\end{equation}
An upwind flux is obtained by adding a dissipative term:
\begin{equation}
 \vect{\mathcal F}_{ij}=\vect f_i+\vect f_j
 -|v^{ij}|(\vect v_i-\vect v_j),
 \qquad
 v^{ij}=\frac12\vect S^{ij}\cdot(\vect v_i+\vect v_j).
 \label{eq:upwind-flux}
\end{equation}
As with all other edge-based upwind fluxes, this first-order scheme 
can be improved by reducing the difference ${\vect v}_i - {\vect v}_j$ through 
(limited) extrapolation to the edge center {\cite{Loh01}}. The same 
scheme is used for the transport equation that describes the 
propagation of the VOF fraction, pseudo-concentration or distance to 
the free surface given by Eqn.~\eqref{eq:transport}. 

\subsubsection{The Divergence Operator}

A persistent difficulty with incompressible flow solvers has been the
derivation of a stable scheme for the divergence 
constraint \eqref{eq:continuity}.
The stability criterion for the divergence constraint is also known as 
the Ladyzenskaya-Babuska-Brezzi or LBB condition \cite{Gun87}. 
The classic way to satisfy the LBB condition has been to use different 
functional spaces for the velocity and pressure discretization 
{\cite{For79}}. Typically, the velocity space has to be
richer, containing more degrees of freedom than the pressure space. 
Elements belonging to this class are the p1/p1+bubble 
mini-element{\cite{Sou87}}, the p1/iso-p1 element {\cite{Tho81}}, and 
the p1/p2 element {\cite{Tay73}}.  An alternative way to satisfy the
LBB condition is through the use of artificial 
viscosities {\cite{Loh90}}, 
`stabilization' {\cite{Fra89, Tez90, Fra92}} or
a `consistent numerical flux'.
The equivalency of these approaches has been repeatedly demonstrated 
{\cite{Sou87, Loh90, Loh01}}.
The approach taken here is based on consistent numerical fluxes, as it
fits naturally into the edge-based framework.
For the divergence constraint, the Galerkin approximation uses the 
central edge flux
\begin{equation}
 \mathcal G_{ij}=\vect S^{ij}\cdot\vect v_i
                 +\vect S^{ij}\cdot\vect v_j.
 \label{eq:central-divergence}
\end{equation}
The equal-order velocity--pressure approximation is stabilized by
adding pressure differences to this flux
\cite{Loh99,Cod01,Lohner2006}. The resulting flux is
\begin{equation}
 \mathcal G_{ij}=\vect S^{ij}\cdot(\vect v_i+\vect v_j)
              -|\lambda^{ij}|(p_i-p_j),
 \qquad
 \lambda^{ij}=\frac{\Delta t^{ij}}{\rho~l^{ij}},
 \label{eq:stabilized-divergence}
\end{equation}
where $l^{ij}=|\vect x_j-\vect x_i|$ is the edge length and
$\Delta t^{ij}$ is the characteristic advective time step for the
edge $\Delta t$. 
Higher order schemes can be derived by reconstruction and limiting,
or by substituting the first-order differences of the pressure with
third-order differences:
\begin{equation}
 \begin{split}
 \mathcal G_{ij}=\vect S^{ij}\cdot(\vect v_i+\vect v_j)
 -|\lambda^{ij}|\Bigl[p_i-p_j+
 \frac12(\vect x_j-\vect x_i)\cdot(\nabla p_i+\nabla p_j)\Bigr].
 \end{split}
 \label{eq:reconstructed-divergence}
\end{equation}
This results in a stable, low-diffusion, fourth-order damping for the 
divergence constraint.

\subsection{Volume of Fluid Extensions}
The extension of a solver for the incompressible Navier-Stokes 
equations to handle free surface flows via the VOF or LS techniques 
requires a series of extensions which are the subject of the present 
section.

\subsubsection{Extrapolation of the Pressure} 
The pressure in the gas region needs to be extrapolated properly in
order to obtain the proper velocities in the region of the free 
surface.
This extrapolation is performed using a three step procedure. In the
first step, the pressures for all point in the gas region are set to
(constant) values, in this case the atmospheric pressure.
In a second step, the gradient of the pressure for the points in the 
liquid that are close to the liquid-gas interface are extrapolated 
from the points inside the liquid region. This step is required as 
the pressure gradient for these points can not be computed properly 
from the data given. Using this information (i.e. pressure and 
gradient of pressure), the pressure for the points in the gas that 
are close to the liquid-gas interface are computed.

\subsubsection{Extrapolation of the Velocity} 
The velocity in the gas region needs to be extrapolated properly in
order to propagate accurately the free surface. This extrapolation
is started by initializing all velocities in the gas region to 
$\vect v=0$.  Then, for each subsequent layer of points in the gas 
region where velocities have not been extrapolated (unknown values), 
an average of the velocities of the surrounding points with known 
values is taken.

\subsubsection{Imposition of Constant Mass} 
Experience indicates that the amount of liquid mass (as measured by 
the region where the VOF indicator is larger than a cut-off value) 
does not remain constant for typical runs. The reasons for this loss 
or gain of mass are manifold: loss of steepness in the interface 
region, inexact divergence of the velocity field, boundary 
velocities, etc. This lack of exact conservation of liquid mass has 
been reported repeatedly in the literature {\cite{Sus94, Sus00, Enr03}}.
The recourse taken here is the classic one: add/remove mass in 
the interface region in order to obtain an exact conservation of 
mass. At the end of every timestep, the total amount of fluid mass 
is compared to the expected value. The expected value is determined 
from the mass at the previous timestep, plus the mass-flux across all 
boundaries during the timestep. The differences in expected and 
actual mass are typically very small, so that
quick convergence is achieved by simply adding and removing mass 
appropriately.  The amount of mass taken/added is made proportional 
to the absolute value of the normal velocity of the interface:
\begin{equation}
 v_n=\left|\vect v\cdot
             \frac{\nabla\Phi}{|\nabla\Phi|}\right|,
 \label{eq:normal-interface-velocity}
\end{equation}
In this way the regions with no movement of the interface remain 
unaffected by the changes made to the interface in order to impose 
strict conservation of mass.

\subsubsection{Deactivation of Air Region} 
Given that the air region is not treated/updated, any CPU spent on it
may be considered wasted. Most of the work is spent in loops over the
edges (upwind solvers, limiters, gradients, etc.). Given that edges 
have to be grouped in order to avoid memory contention/ allow 
vectorization when forming right-hand sides {\cite{Loh93, Loh98}}, 
this opens a natural way of avoiding unnecessary work: form 
relatively small edge-groups that 
still allow for efficient vectorization, and deactivate groups instead 
of individual edges {\cite{Loh01}}. In this way, the basic loops 
over edges do not require any changes.
The {\code if}-test whether an edge group is active or deactive occurs
outside the inner loops over edges, leaving them unaffected.
On scalar processors, edges-groups as small as {\code negrp=8} are
used. Furthermore, if points and edges are grouped together in such a 
way that proximity in memory mirrors spatial proximity, most of the 
edges in air will not incur any CPU penalty.

\section{Computational Model}
\label{sec:computational-model}

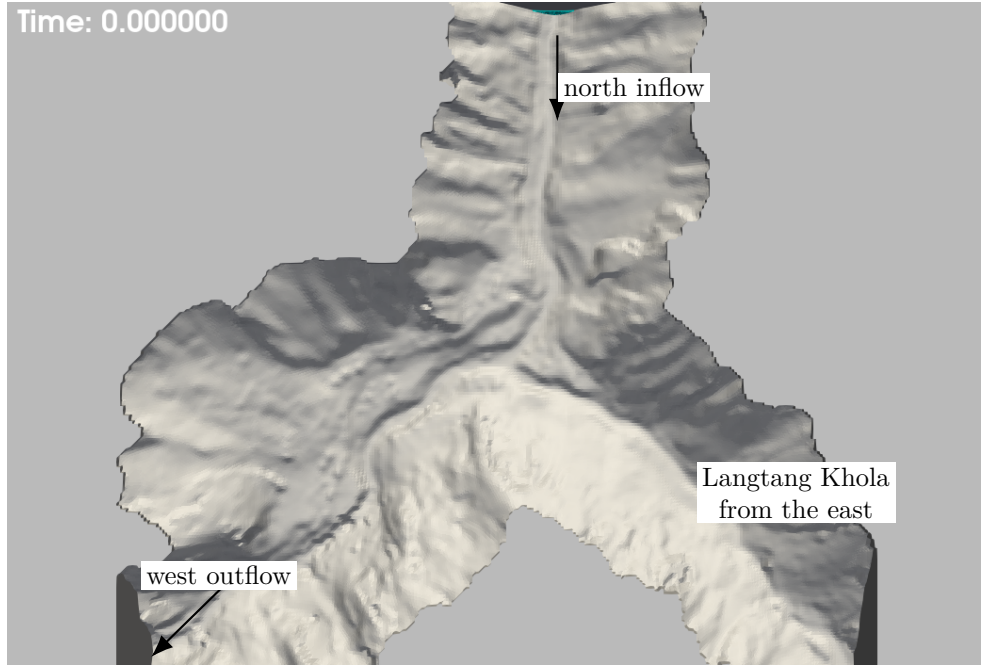
\begin{figure}[htbp]
\centering
\begin{tikzpicture}
\node[anchor=south west,inner sep=0] (img) at (0,0)
{\includegraphics[width=0.78\linewidth]{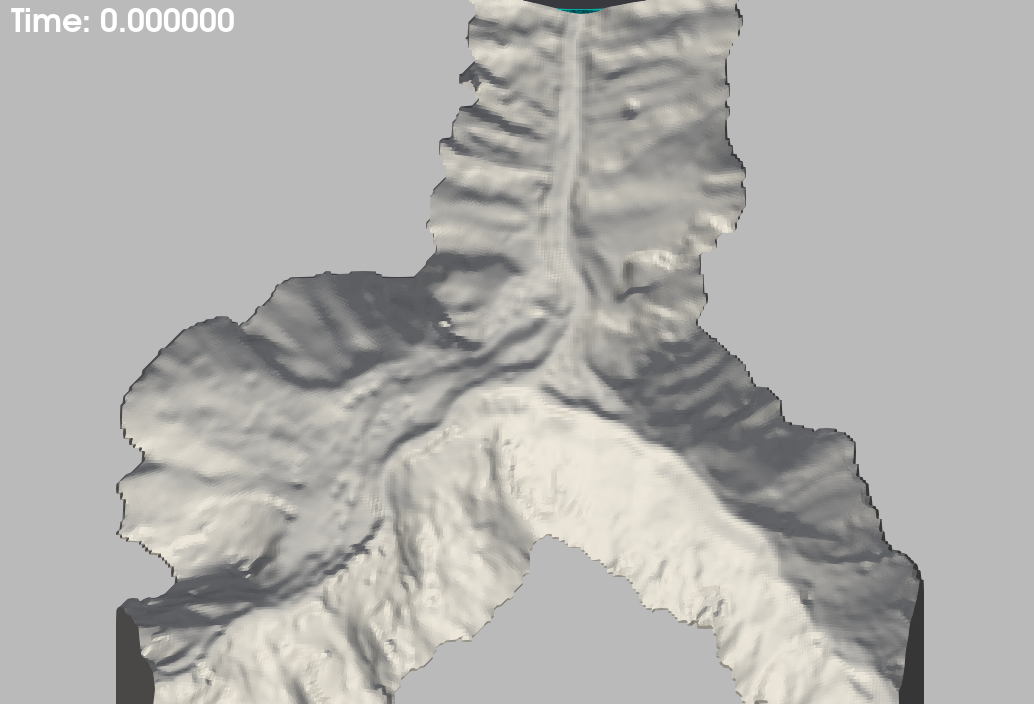}};
\begin{scope}[x={(img.south east)},y={(img.north west)}]
\draw[-{Latex[length=3mm]},thick]
 (0.565,0.95)--(0.565,0.82)
 node[pos=0.60,right,fill=white,inner sep=2pt,font=\small]{north inflow};
\draw[-{Latex[length=3mm]},thick]
 (0.23,0.13)--(0.148,0.012)
 node[pos=0.15,above,fill=white,inner sep=2pt,font=\small]{west outflow};
\node[fill=white,inner sep=2pt,font=\small,align=center]
 at (0.81,0.26) {Langtang Khola\\from the east};
\end{scope}
\end{tikzpicture}
\caption{Computational terrain at Syabrubesi. The flood enters along the
Bhote Koshi from the north, turns southwest through the junction and
leaves through the western river section.}
\label{fig:domain}
\end{figure}

The terrain covers a $2\,\mathrm{km}\times2\,\mathrm{km}$ square
centred near the Bhote Koshi--Langtang Khola junction and is sampled
on a pre-event $8\,\mathrm{m}$ grid. The northern boundary intersects
the Bhote Koshi upstream of the junction. The river turns southwest
and leaves through the western boundary; the Langtang Khola joins
from the east. Figure~\ref{fig:domain} shows the computational terrain
and the direction of the imposed flood.

The surface triangulation was imported into the in-house preprocessor
FECAD. This allowed the higher elevations to be cut, the domain to be
closed, the boundary regions to be assigned and the desired element-size
distribution to be specified. An advancing-front grid generator was
then used to mesh the surface and volume
\cite{LohnerParikh1988,Lohner1996Mesh,Lohner2001Mesh}. Refinement was
concentrated near river level to resolve the narrow channel and the
free-surface motion.

At the northern boundary, the inflow speed was set to
$v_{\mathrm{in}}=10\,\mathrm{m\,s^{-1}}$ and the maximum water depth
was approximately $40\,\mathrm{m}$. Given the uncertainty in the
composition of the mud, the density was assumed to be that of water,
$\rho=1000\,\mathrm{kg\,m^{-3}}$. The original river flow was neglected
relative to the flood. The runs were carried out to $180\,\mathrm{s}$.

Meshes with approximately 5, 20 and 100 million elements were used to
examine changes in the peak velocities and depths under refinement.
A preliminary mesh comparison showed no marked
differences in these quantities for the 20 and 100 million element
grids. The results presented below were
obtained on the mesh with approximately 100 million elements.

\begin{figure}[htbp]
\centering
\begin{subfigure}[t]{0.62\linewidth}
\centering
\includegraphics[width=\linewidth]{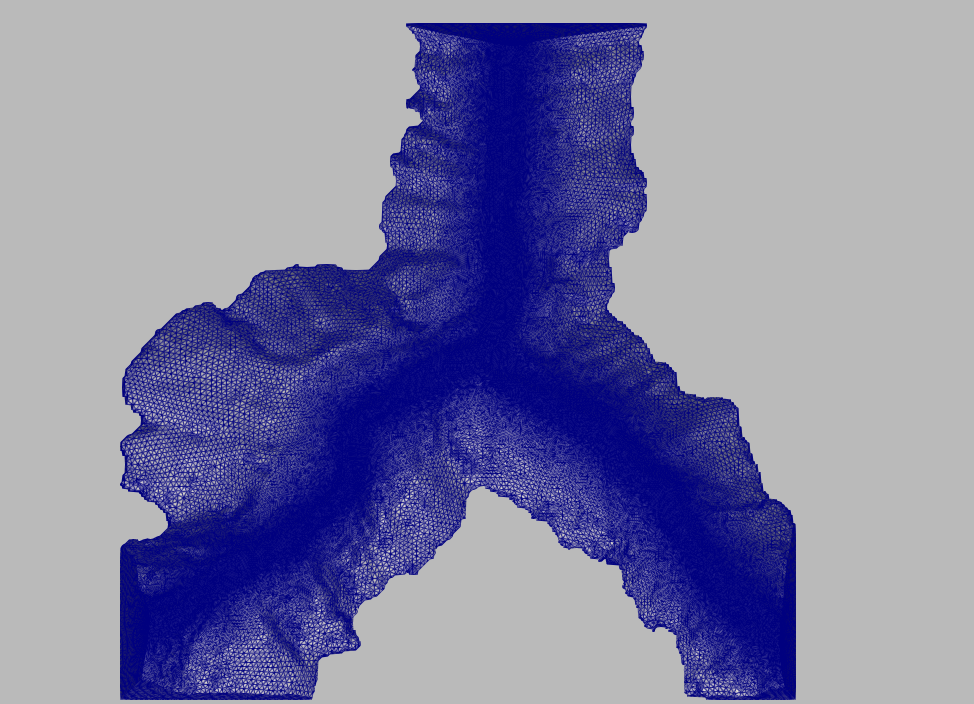}
\caption{Overall view}
\end{subfigure}
\par\medskip
\begin{subfigure}[t]{0.485\linewidth}\centering
  \includegraphics[width=\linewidth]{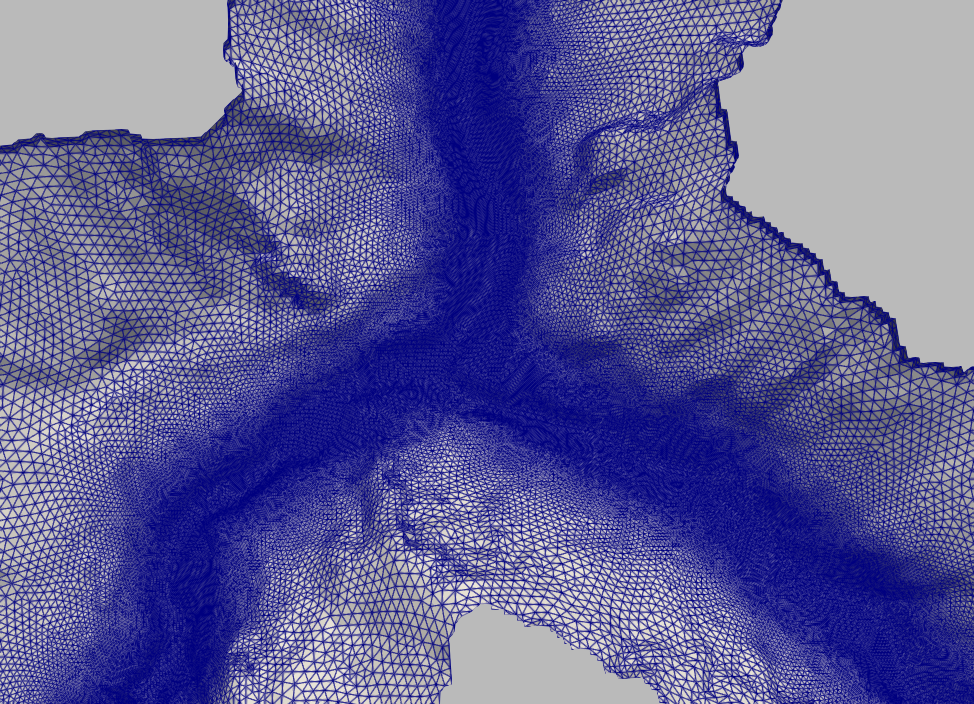}\caption{Junction}
  \end{subfigure}\hfill
\begin{subfigure}[t]{0.485\linewidth}\centering
  \includegraphics[width=\linewidth]{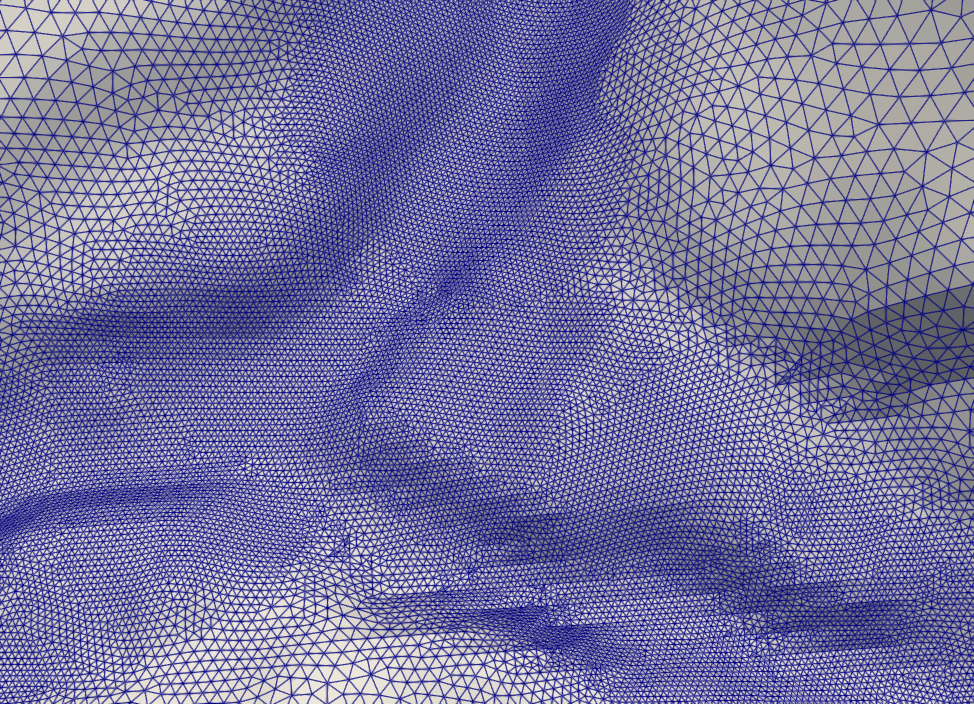}\caption{Detail of the channel refinement}
  \end{subfigure}
\caption{Surface triangulation of the computational domain, showing
refinement near the river channel.}
\label{fig:mesh}
\end{figure}

Figure~\ref{fig:mesh} shows the surface triangulation in an
overall view and two close-ups near the junction. The smaller elements
follow the river channel and transition to larger elements on the
surrounding slopes. The edge lengts for the elements close to the
river bed are of the order of 1.25~m.

\section{Results}
\label{sec:results}

\begin{figure}[htbp]
\centering
\begin{subfigure}[t]{0.485\linewidth}\centering
  \includegraphics[width=\linewidth]{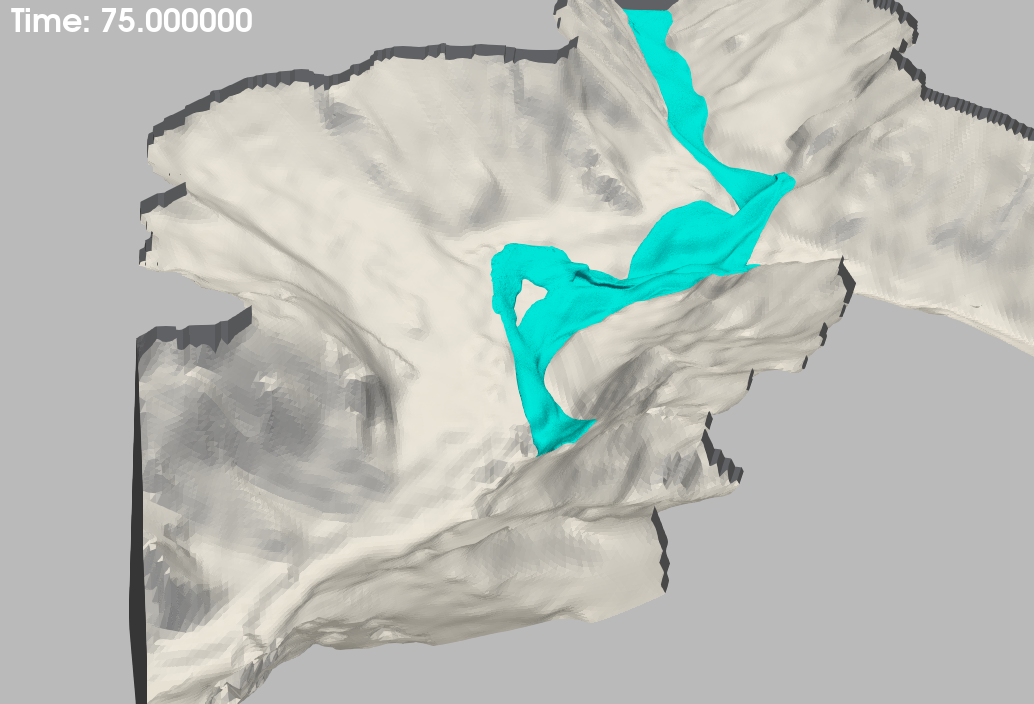}\caption{$t=75\,\mathrm{s}$}
  \end{subfigure}\hfill
\begin{subfigure}[t]{0.485\linewidth}\centering
  \includegraphics[width=\linewidth]{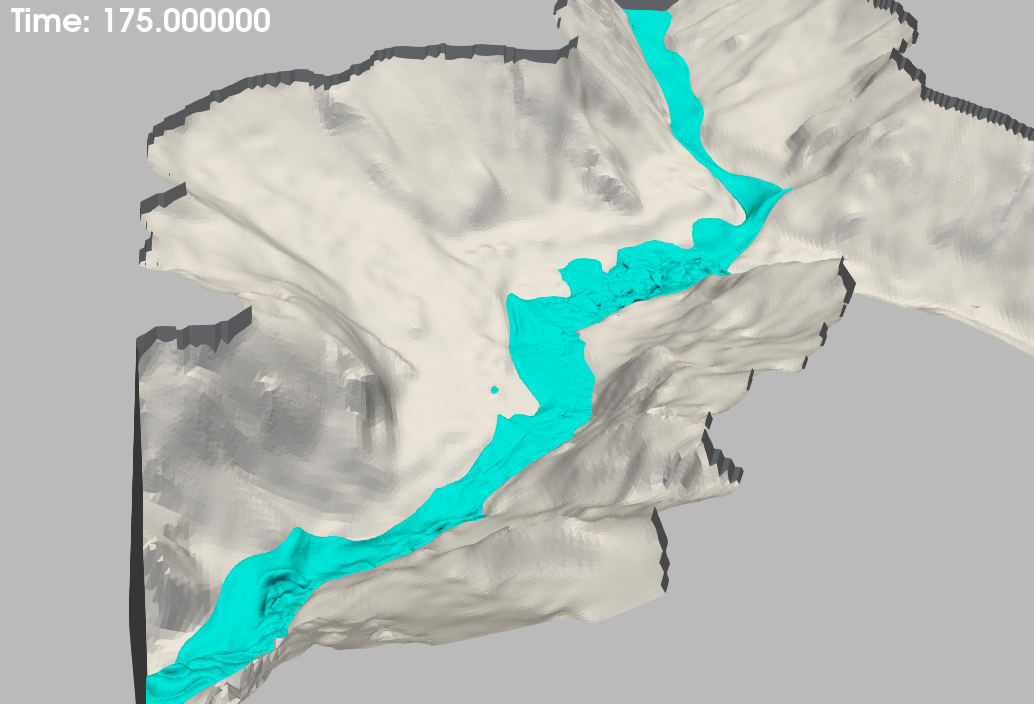}\caption{$t=175\,\mathrm{s}$}
  \end{subfigure}
\caption{Free-surface evolution through the junction, shown from a second
viewing angle.}
\label{fig:surface}
\end{figure}

\begin{figure}[htbp]
\centering
\begin{subfigure}[t]{0.485\linewidth}\centering
  \includegraphics[width=\linewidth]{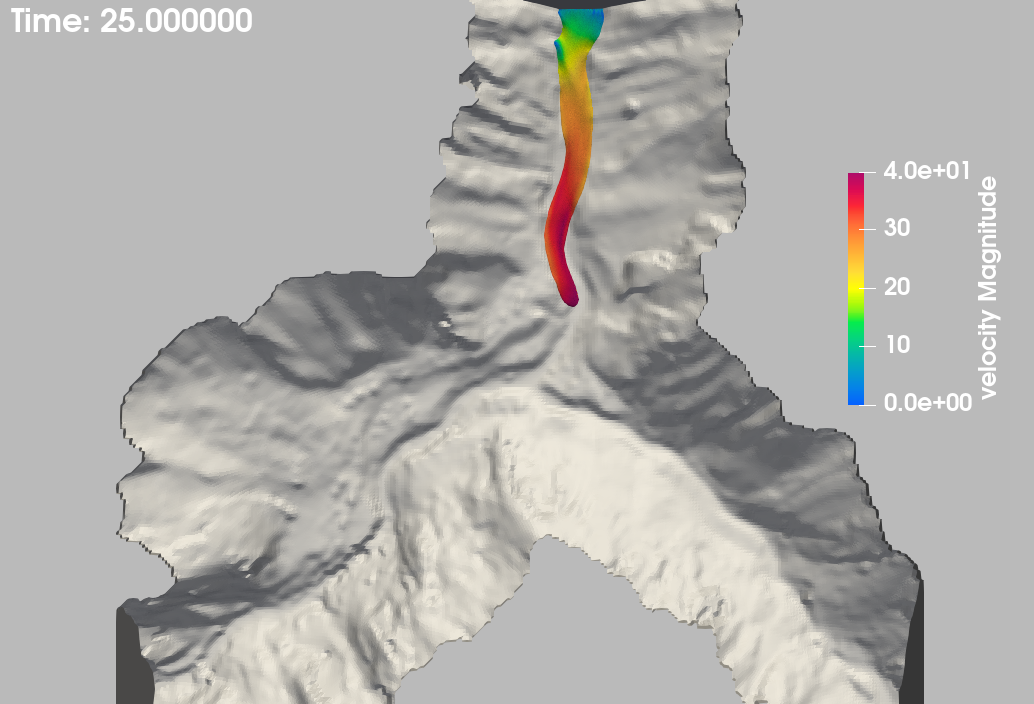}\caption{$t=25\,\mathrm{s}$}
  \end{subfigure}\hfill
\begin{subfigure}[t]{0.485\linewidth}\centering
  \includegraphics[width=\linewidth]{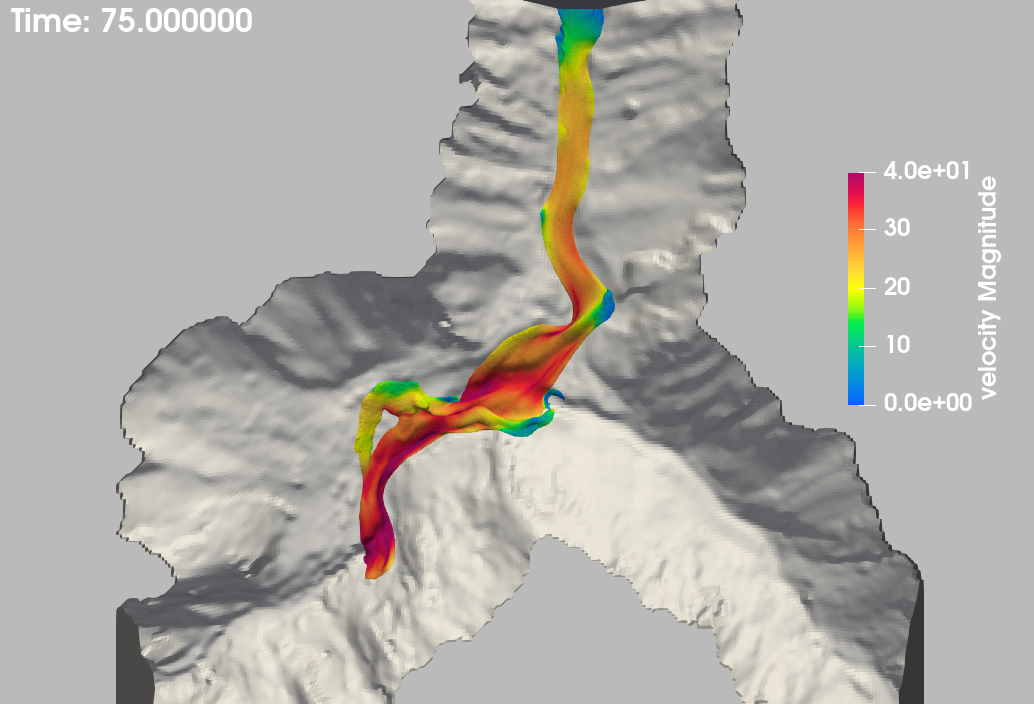}\caption{$t=75\,\mathrm{s}$}
  \end{subfigure}
\par\medskip
\begin{subfigure}[t]{0.485\linewidth}\centering
  \includegraphics[width=\linewidth]{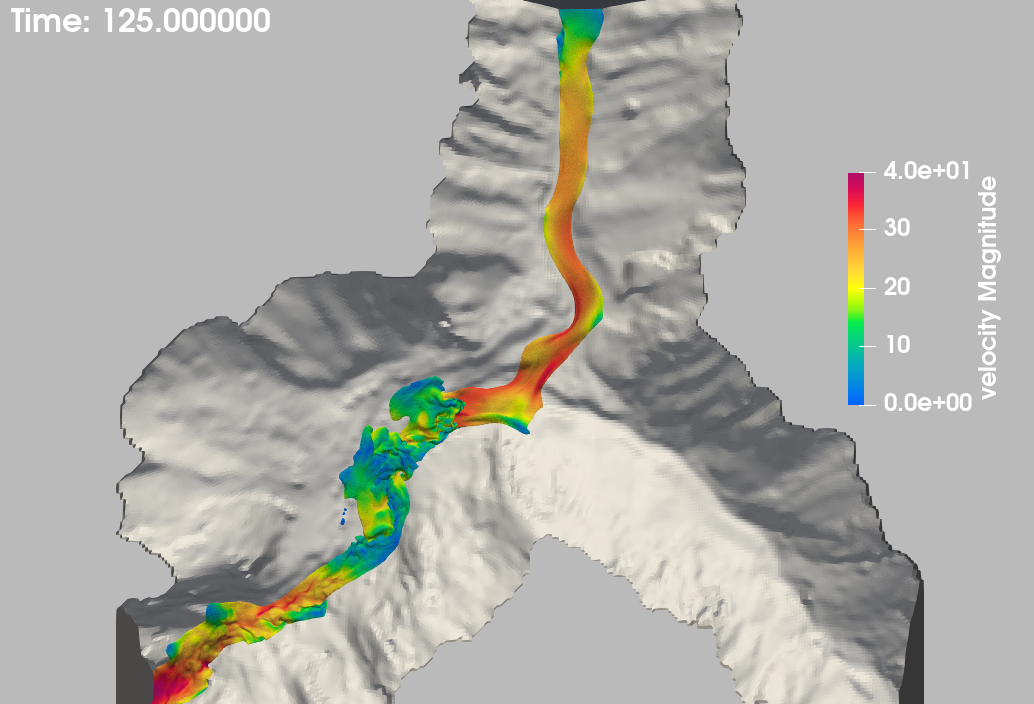}\caption{$t=125\,\mathrm{s}$}
  \end{subfigure}\hfill
\begin{subfigure}[t]{0.485\linewidth}\centering
  \includegraphics[width=\linewidth]{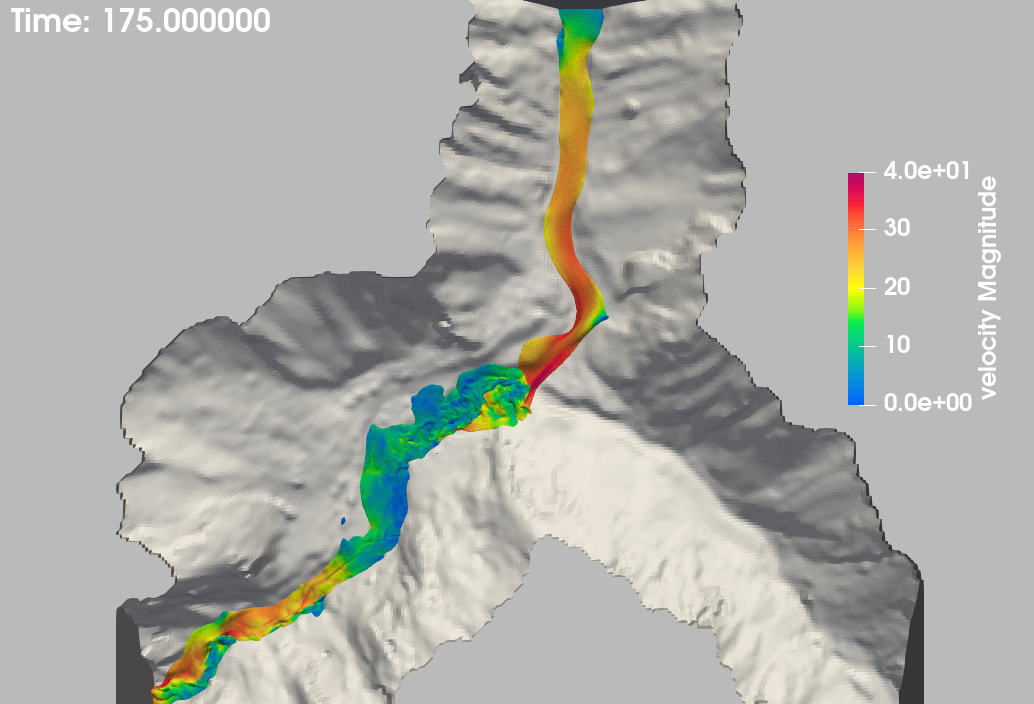}\caption{$t=175\,\mathrm{s}$}
  \end{subfigure}
\caption{Velocity magnitude on the free surface, in
$\mathrm{m\,s^{-1}}$. The same $0$--$40$ colour scale is used at all
four times.}
\label{fig:velocity}
\end{figure}

Figures~\ref{fig:surface} and \ref{fig:velocity} show the passage of
the flood through the junction. At $25\,\mathrm{s}$ the advancing
flood is still in the confined northern reach. It has crossed the
junction and rounded the bend by $75\,\mathrm{s}$, and reaches the
western outlet by $125\,\mathrm{s}$. The flow spreads laterally as
the channel widens at the junction and then follows the narrower
downstream reach. The oblique views in Figure~\ref{fig:surface}
show the irregular free surface around the bend at $75$ and
$175\,\mathrm{s}$.

The velocity increases from the prescribed inlet value of
$10\,\mathrm{m\,s^{-1}}$ as the flood descends the incoming channel.
Surface speeds above $30\,\mathrm{m\,s^{-1}}$ occur in parts of this
confined reach (Figure~\ref{fig:velocity}). A fast core follows the
channel, with lower velocities along its margins. At $75\,\mathrm{s}$
the advancing flow remains fast as it crosses the bend. By $125$ and
$175\,\mathrm{s}$ a broader region of slower flow has formed there,
while higher surface speeds occur again in the confined downstream
channel.

\begin{figure}[htbp]
\centering
\begin{subfigure}[t]{0.485\linewidth}\centering
  \includegraphics[width=\linewidth]{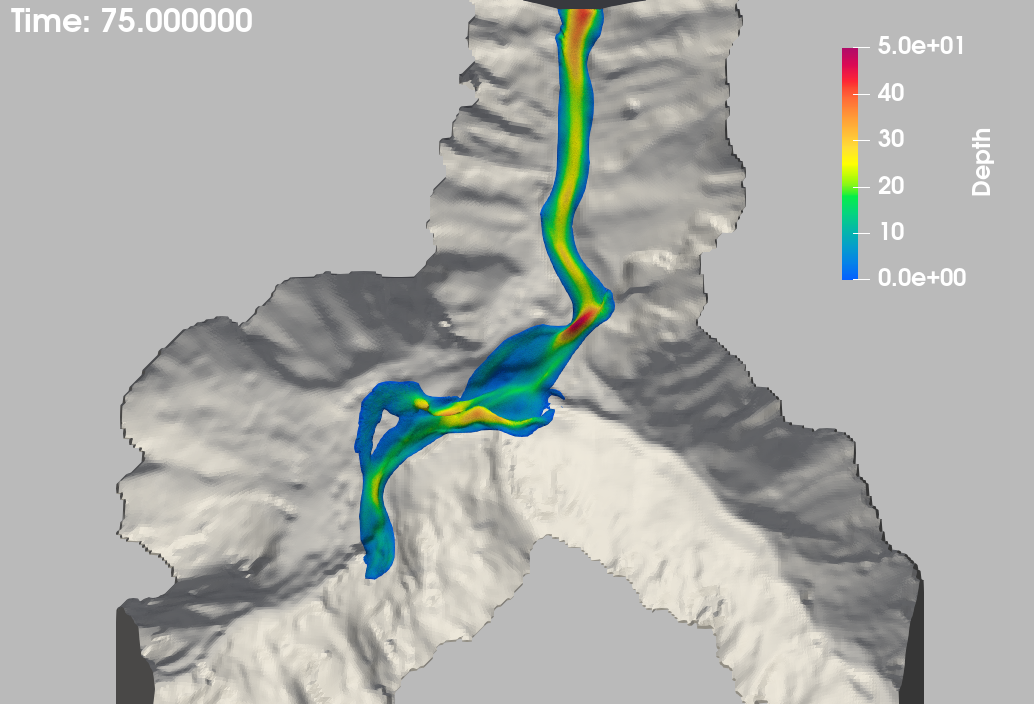}\caption{Depth, $t=75\,\mathrm{s}$}
  \end{subfigure}\hfill
\begin{subfigure}[t]{0.485\linewidth}\centering
  \includegraphics[width=\linewidth]{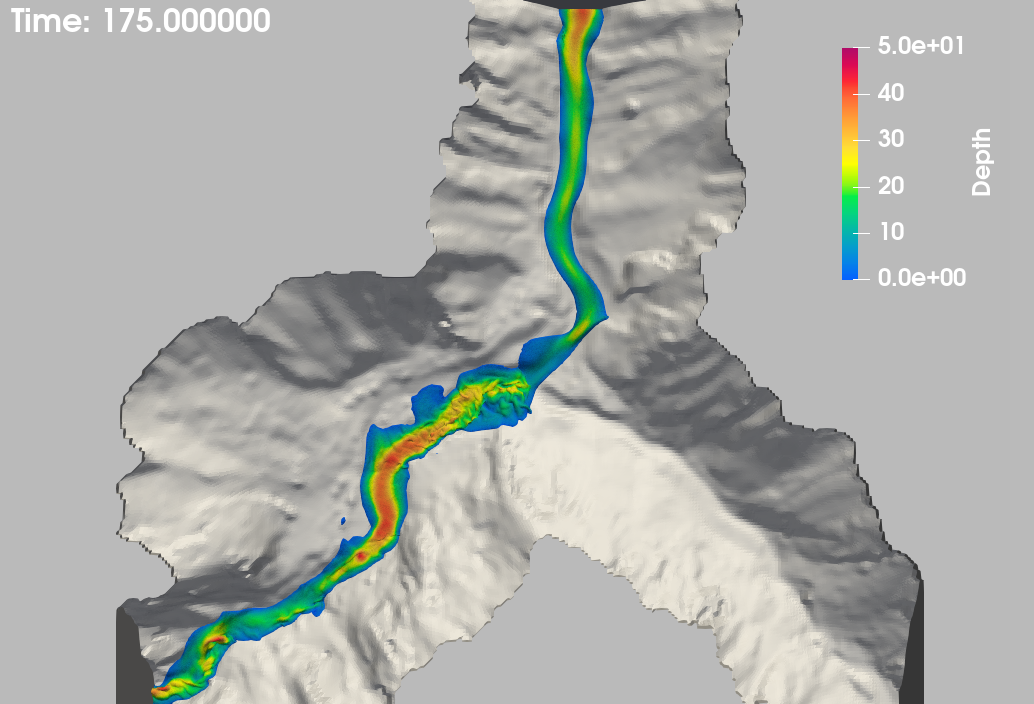}\caption{Depth, $t=175\,\mathrm{s}$}
  \end{subfigure}
\par\medskip
\begin{subfigure}[t]{0.485\linewidth}\centering
  \includegraphics[width=\linewidth]{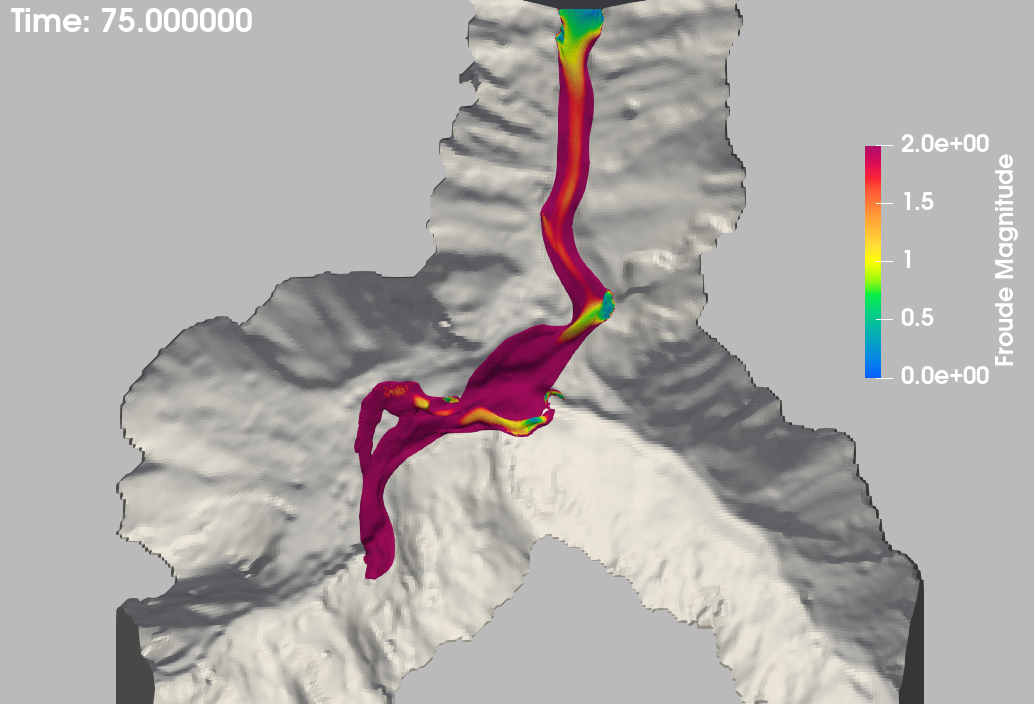}\caption{Froude number, $t=75\,\mathrm{s}$}
  \end{subfigure}\hfill
\begin{subfigure}[t]{0.485\linewidth}\centering
  \includegraphics[width=\linewidth]{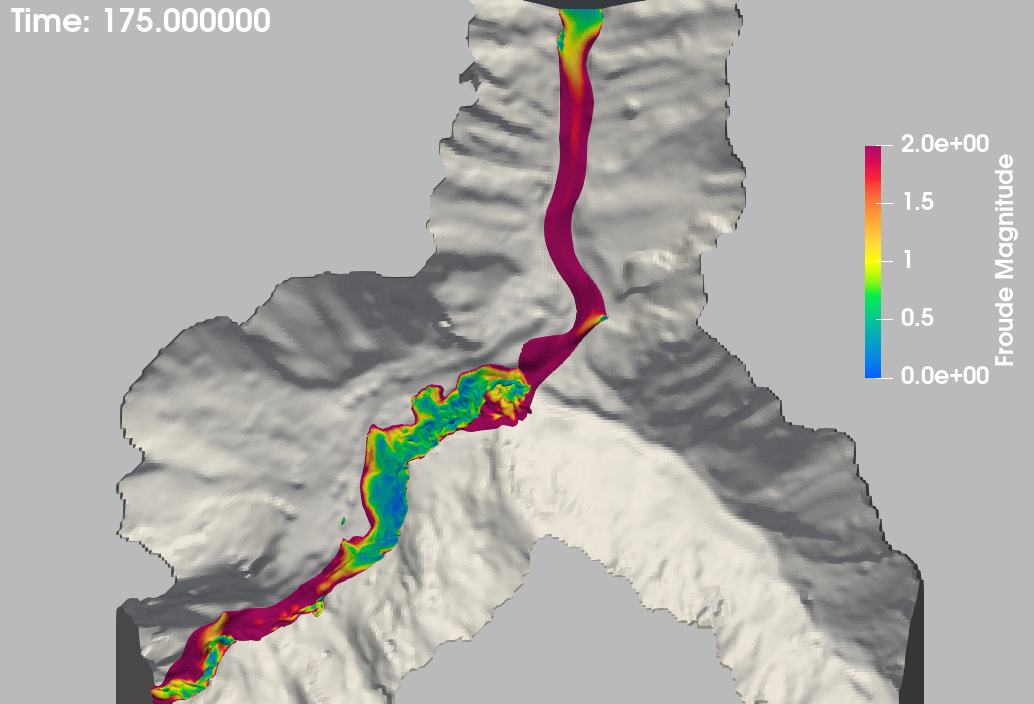}\caption{Froude number, $t=175\,\mathrm{s}$}
  \end{subfigure}
\caption{Water depth (top, in metres) and Froude number (bottom) at
$75$ and $175\,\mathrm{s}$. The depth scale is $0$--$50\,\mathrm{m}$
and the Froude scale is $0$--$2$.}
\label{fig:depth-froude}
\end{figure}

The change in velocity around the bend is accompanied by an increase
in water depth. Figure~\ref{fig:depth-froude} compares the depth at
$75$ and $175\,\mathrm{s}$. The advancing front crosses the bend
with relatively shallow water, whereas deeper water occupies much of
the bend at $175\,\mathrm{s}$. The depth is nonuniform across this region,
with shallower water at the margins. Thus the bend has a different
transient response from the faster, confined reaches on either side.

\begin{samepage}
The corresponding Froude number, defined by
\begin{equation}
 \mathrm{Fr}=\frac{|\vect v|}{\sqrt{gH}},
 \label{eq:froude}
\end{equation}
where $H$ is the total depth of the water column and $g=|\vect g|$
is a useful measure to guage if the flow is supercritical and thus
if information can travel upstream.
Much of the confined channel has $\mathrm{Fr}>1$, i.e. is 
supercritical.
At $75\,\mathrm{s}$, high values extend across most of the advancing
flow at the bend. By $175\,\mathrm{s}$, lower values occupy parts of
the deeper, slower flow in this region, while high values remain in
the confined upstream and downstream reaches.
\end{samepage}

\section{Conclusions}

Three-dimensional free-surface computations have been carried out for
flood propagation through the Bhote Koshi--Langtang Khola junction at
Syabrubesi. The terrain was represented by unstructured tetrahedral
meshes with refinement near the river, and the flow was computed
using the FEFLO code. The
calculations show a fast front spreading through the junction,
followed by deeper and slower flow around the bend. Confined reaches
upstream and downstream retain higher velocities, with surface speeds
above $40\,\mathrm{m\,s^{-1}}$ in parts of the incoming channel for
the prescribed $10\,\mathrm{m\,s^{-1}}$ inflow. We remark in passing that
velocities of this order of magnitude were actually observed in the 
flooding event.

These results describe the local hydraulic response under the assumed
inflow. The terrain predates the event,
does not resolve river bathymetry and is held fixed during the
calculation. Sediment transport and changes in bed elevation are not
included.

\bibliographystyle{aiaa}

\end{document}